\documentstyle[12pt]{article}

\author{
{\normalsize James Botts$^a$, Jorge G. Morfin$^b$, Joseph F. Owens$^c$,
Jianwei Qiu$^d$, }\\
{\normalsize  Wu-Ki Tung$^a$ and Harry Weerts$^a$}\\
\rule{0mm}{1mm}\\
{\normalsize $^a$Michigan State University, }\\
{\normalsize $^b$Fermi National Accelerator Laboratory, }\\
{\normalsize $^c$Florida State University, $^d$Iowa State University}\\
\rule{0in}{1mm} \\
{\normalsize (to be published in Physics Letters)}
}
\title{
CTEQ Parton Distributions and Flavor Dependence of Sea Quarks\thanks{The CTEQ
Collaboration is funded by Texas National Laboratory Commission.
This work is also partially supported by NSF and DOE through grants made to
the home institutions of the authors.}
}
\date{\null
}

\newcommand{\Ref}[1]{{\small \cite{#1}}}

\newcommand{\lo} {{\protect \footnotesize LO}}

\newcommand{\dis}{{\protect \footnotesize DIS}}

\newcommand{\Msbar}{\mbox{\footnotesize {$\overline {\rm MS}$}}}

\catcode`\@=11

\let\DOTSI\relax
\def\RIfM@{\relax\ifmmode}
\def\FN@{\futurelet\next}
\newcount\intno@
\def\iint{\DOTSI\intno@\tw@\FN@\ints@}
\def\iiint{\DOTSI\intno@\thr@@\FN@\ints@}
\def\iiiint{\DOTSI\intno@4 \FN@\ints@}
\def\idotsint{\DOTSI\intno@\z@\FN@\ints@}
\def\ints@{\findlimits@\ints@@}
\newif\iflimtoken@
\newif\iflimits@
\def\findlimits@{\limtoken@true\ifx\next\limits\limits@true
 \else\ifx\next\nolimits\limits@false\else
 \limtoken@false\ifx\ilimits@\nolimits\limits@false\else
 \ifinner\limits@false\else\limits@true\fi\fi\fi\fi}
\def\multint@{\int\ifnum\intno@=\z@\intdots@                                
 \else\intkern@\fi                                                          
 \ifnum\intno@>\tw@\int\intkern@\fi                                         
 \ifnum\intno@>\thr@@\int\intkern@\fi                                       
 \int}                                                                      
\def\multintlimits@{\intop\ifnum\intno@=\z@\intdots@\else\intkern@\fi
 \ifnum\intno@>\tw@\intop\intkern@\fi
 \ifnum\intno@>\thr@@\intop\intkern@\fi\intop}
\def\intic@{\mathchoice{\hskip.5em}{\hskip.4em}{\hskip.4em}{\hskip.4em}}
\def\negintic@{\mathchoice
 {\hskip-.5em}{\hskip-.4em}{\hskip-.4em}{\hskip-.4em}}
\def\ints@@{\iflimtoken@                                                    
 \def\ints@@@{\iflimits@\negintic@\mathop{\intic@\multintlimits@}\limits    
  \else\multint@\nolimits\fi                                                
  \eat@}                                                                    
 \else                                                                      
 \def\ints@@@{\iflimits@\negintic@
  \mathop{\intic@\multintlimits@}\limits\else
  \multint@\nolimits\fi}\fi\ints@@@}
\def\intkern@{\mathchoice{\!\!\!}{\!\!}{\!\!}{\!\!}}
\def\plaincdots@{\mathinner{\cdotp\cdotp\cdotp}}
\def\intdots@{\mathchoice{\plaincdots@}
 {{\cdotp}\mkern1.5mu{\cdotp}\mkern1.5mu{\cdotp}}
 {{\cdotp}\mkern1mu{\cdotp}\mkern1mu{\cdotp}}
 {{\cdotp}\mkern1mu{\cdotp}\mkern1mu{\cdotp}}}

\newif\iffirstchoice@
\firstchoice@true
\def\textfonti{\the\textfont\@ne}
\def\textfontii{\the\textfont\tw@}
\def\text{\RIfM@\expandafter\text@\else\expandafter\text@@\fi}
\def\text@@#1{\leavevmode\hbox{#1}}
\def\text@#1{\mathchoice
 {\hbox{\everymath{\displaystyle}\def\textfonti{\the\textfont\@ne}%
  \def\textfontii{\the\textfont\tw@}\textdef@@ T#1}}
 {\hbox{\firstchoice@false
  \everymath{\textstyle}\def\textfonti{\the\textfont\@ne}%
  \def\textfontii{\the\textfont\tw@}\textdef@@ T#1}}
 {\hbox{\firstchoice@false
  \everymath{\scriptstyle}\def\textfonti{\the\scriptfont\@ne}%
  \def\textfontii{\the\scriptfont\tw@}\textdef@@ S\rm#1}}
 {\hbox{\firstchoice@false
  \everymath{\scriptscriptstyle}\def\textfonti
  {\the\scriptscriptfont\@ne}%
  \def\textfontii{\the\scriptscriptfont\tw@}\textdef@@ s\rm#1}}}
\def\textdef@@#1{\textdef@#1\rm\textdef@#1\bf\textdef@#1\sl\textdef@#1\it}
\def\DN@{\def\next@}
\def\eat@#1{}
\def\textdef@#1#2{%
 \DN@{\csname\expandafter\eat@\string#2fam\endcsname}%
 \if S#1\edef#2{\the\scriptfont\next@\relax}%
 \else\if s#1\edef#2{\the\scriptscriptfont\next@\relax}%
 \else\edef#2{\the\textfont\next@\relax}\fi\fi}

\def\Let@{\relax\iffalse{\fi\let\\=\cr\iffalse}\fi}
\def\vspace@{\def\vspace##1{\crcr\noalign{\vskip##1\relax}}}
\def\multilimits@{\bgroup\vspace@\Let@
 \baselineskip\fontdimen10 \scriptfont\tw@
 \advance\baselineskip\fontdimen12 \scriptfont\tw@
 \lineskip\thr@@\fontdimen8 \scriptfont\thr@@
 \lineskiplimit\lineskip
 \vbox\bgroup\ialign\bgroup\hfil$\m@th\scriptstyle{##}$\hfil\crcr}
\def\Sb{_\multilimits@}
\def\endSb{\crcr\egroup\egroup\egroup}
\def\Sp{^\multilimits@}

\newdimen\ex@
\def\rightarrowfill@#1{$#1\m@th\mathord-\mkern-6mu\cleaders
 \hbox{$#1\mkern-2mu\mathord-\mkern-2mu$}\hfill
 \mkern-6mu\mathord\rightarrow$}
\def\leftarrowfill@#1{$#1\m@th\mathord\leftarrow\mkern-6mu\cleaders
 \hbox{$#1\mkern-2mu\mathord-\mkern-2mu$}\hfill\mkern-6mu\mathord-$}
\def\leftrightarrowfill@#1{$#1\m@th\mathord\leftarrow\mkern-6mu\cleaders
 \hbox{$#1\mkern-2mu\mathord-\mkern-2mu$}\hfill
 \mkern-6mu\mathord\rightarrow$}
\def\overrightarrow{\mathpalette\overrightarrow@}
\def\overrightarrow@#1#2{\vbox{\ialign{##\crcr\rightarrowfill@#1\crcr
 \noalign{\kern-\ex@\nointerlineskip}$\m@th\hfil#1#2\hfil$\crcr}}}

\def\overleftarrow{\mathpalette\overleftarrow@}
\def\overleftarrow@#1#2{\vbox{\ialign{##\crcr\leftarrowfill@#1\crcr
 \noalign{\kern-\ex@\nointerlineskip}$\m@th\hfil#1#2\hfil$\crcr}}}
\def\overleftrightarrow{\mathpalette\overleftrightarrow@}
\def\overleftrightarrow@#1#2{\vbox{\ialign{##\crcr\leftrightarrowfill@#1\crcr
 \noalign{\kern-\ex@\nointerlineskip}$\m@th\hfil#1#2\hfil$\crcr}}}
\def\underrightarrow{\mathpalette\underrightarrow@}
\def\underrightarrow@#1#2{\vtop{\ialign{##\crcr$\m@th\hfil#1#2\hfil$\crcr
 \noalign{\nointerlineskip}\rightarrowfill@#1\crcr}}}

\def\underleftarrow{\mathpalette\underleftarrow@}
\def\underleftarrow@#1#2{\vtop{\ialign{##\crcr$\m@th\hfil#1#2\hfil$\crcr
 \noalign{\nointerlineskip}\leftarrowfill@#1\crcr}}}
\def\underleftrightarrow{\mathpalette\underleftrightarrow@}
\def\underleftrightarrow@#1#2{\vtop{\ialign{##\crcr$\m@th\hfil#1#2\hfil$\crcr
 \noalign{\nointerlineskip}\leftrightarrowfill@#1\crcr}}}

\catcode`\@=\active

\def\frac#1#2{{#1 \over #2}}

\newcount\GRAPHICSTYPE
\def\GRAPHICSPS#1{%
\ifnum\GRAPHICSTYPE=1 language "PS", include "#1"\else%
ps: #1\fi}

\def\graffile#1#2#3#4{\leavevmode\raise -#4 \hbox{%
\raise #3 \hbox{\rule{0.003in}{0.003in}\special{#1}}}%
{\raise -#4 \hbox to #2 {\vrule height#3 width0in depth0in\hfil}}%
}

\def\draftbox#1#2#3#4{\leavevmode\raise -#4 \hbox{\frame{\rlap{\protect\tiny
#1}%
\hbox to #2{\vrule height#3 width0in depth0in\hfil}}}}

\newcount\draft
\def\GRAPHIC#1#2#3#4#5{\ifnum\draft=1 \draftbox{#2}{#3}{#4}{#5}\else%
\graffile{#1}{#3}{#4}{#5}\fi}

\def\addtoLaTeXparams#1{\edef\LaTeXparams{\LaTeXparams #1}}

\def\doFRAMEparams#1{\readFRAMEparams#1\end}
\def\readFRAMEparams#1{%
\ifx#1\end%
\let\next=\relax%
\else%
\ifx#1i%
\dispkind=0%
\fi%
\ifx#1d%
\dispkind=1%
\fi%
\ifx#1f%
\dispkind=2%
\fi%
\ifx#1t%
\addtoLaTeXparams{t}%
\fi%
\ifx#1b%
\addtoLaTeXparams{b}%
\fi%
\ifx#1p%
\addtoLaTeXparams{p}%
\fi%
\ifx#1h%
\addtoLaTeXparams{h}%
\fi%
\let\next=\readFRAMEparams%
\fi%
\next%
}

\def\IFRAME#1#2#3#4#5{\GRAPHIC{#5}{#4}{#1}{#2}{#3}}

\def\DFRAME#1#2#3#4{
  \begin{center}
    \GRAPHIC{#4}{#3}{#1}{#2}{0in}
  \end{center}
}

\def\FFRAME#1#2#3#4#5#6#7{
  \begin{figure}[#1]
    \begin{center}
      \GRAPHIC{#7}{#6}{#2}{#3}{0in}
    \end{center}
    \caption{\label{#5}#4}
  \end{figure}
}

\def\FRAME#1#2#3#4#5#6#7#8{%
\newcount\dispkind%
\def\LaTeXparams{}%
\dispkind=0%
\def\LaTeXparams{}%
\doFRAMEparams{#1}%
\ifnum\dispkind=0%
\IFRAME{#2}{#3}{#4}{#7}{#8}%
\else
  \ifnum\dispkind=1
    \DFRAME{#2}{#3}{#7}{#8}
  \else
    \ifnum\dispkind=2
      \FFRAME{\LaTeXparams}{#2}{#3}{#5}{#6}{#7}{#8}
    \fi
  \fi
\fi
}

\catcode`\@=11

\long\def\QQQ#1#2{}
\def\QTP#1{}
\long\def\QQA#1#2{}

\def\EXPAND#1[#2]#3{}
\def\NOEXPAND#1[#2]#3{}

\def\LaTeXparent#1{}

\@ifundefined{INDEX}{\def\INDEX#1#2{}{}}{}
\@ifundefined{SUBINDEX}{\def\SUBINDEX#1#2#3{}{}{}}{}
\def\initial#1{\bigbreak{\raggedright\large\bf #1}\kern 2pt\penalty3000}

\@ifundefined{abstract}{%
\def\abstract{\if@twocolumn
\section*{Abstract (Not appropriate in this style!)}
\else \small
\begin{center}
{\bf Abstract\vspace{-.5em}\vspace{0pt}}
\end{center}
\quotation
\fi}}{}
\@ifundefined{endabstract}{%
\def\endabstract{\if@twocolumn\else\endquotation\fi}}{}
\@ifundefined{maketitle}{\def\maketitle#1{}}{}
\@ifundefined{affiliation}{\def\affiliation#1{}}{}
\@ifundefined{proof}{}{}
\@ifundefined{newfield}{\def\newfield#1#2{}}{}
\@ifundefined{chapter}{\def\chapter#1{\par(Chapter head:)#1\par }}{}
\@ifundefined{part}{\def\part#1{\par(Part head:)#1\par }}{}
\@ifundefined{section}{\def\part#1{\par(Section head:)#1\par }}{}
\@ifundefined{subsection}{\def\part#1{\par(Subsection head:)#1\par }}{}
\@ifundefined{subsubsection}{\def\part#1{\par(Subsubsection head:)#1\par }}{}
\@ifundefined{paragraph}{\def\part#1{\par(Subsubsubsection head:)#1\par }}{}
\@ifundefined{subparagraph}{\def\part#1{\par(Subsubsubsubsection head:)#1\par
}}{}

\newdimen\theight
\def \Column{%
             \vadjust{\setbox0=\hbox{\scriptsize\quad\quad tcol}%
             \theight=\ht0
             \advance\theight by \dp0    \advance\theight by \lineskip
             \kern -\theight \vbox to \theight{\rightline{\rlap{\box0}}%
             \vss}%
             }}%

\def\qed{\ifhmode\unskip\nobreak\fi\ifmmode\ifinner\else\hskip5\p@\fi\fi
 \hbox{\hskip5\p@\vrule width4\p@ height6\p@ depth1.5\p@\hskip\p@}}
\catcode`@=12 

\makeatletter

\begin{document}

\maketitle
\begin{abstract}
{This paper describes salient features of new sets of parton distributions
obtained by the CTEQ Collaboration%
\footnote{CTEQ is an acronym for Coordinated Theoretical/Experimental Project
on QCD
Phenomenology and Tests of the Standard Model. The Collaboration consists of,
in addition to the above
authors (as members of its global fit subgroup), R. Brock, J. Huston, J.
Pumplin,
C.P. Yuan (MSU); J. Collins, J. Whitmore (PSU); S. Kuhlmann (Argonne);
S. Mishra (Harvard); F. Olness (SMU); D. Soper (Oregon);
J. Smith, and G. Sterman (Stony Brook).} based on a comprehensive QCD global
analysis of all available data. The accuracy of the new data on deep
inelastic scattering structure functions obtained by the very high
statistics NMC and CCFR experiments provides unprecedented sensitivity to
the flavor dependence of the sea-quark distributions. In addition to much
better determination of the small }$x${\ dependence of all parton
distributions, we found: (i) the strange quark distribution is much softer
than the non-strange sea quarks and rises above the latter at small-$x$; and
(ii) the difference }$\bar d-\bar u$ changes sign as a function of $x${. A
few alternative sets of viable distributions with conventional assumptions
are also discussed.}
\end{abstract}

\pagebreak
\baselineskip=0.23in %
\pagestyle{plain}
\paragraph{Introduction}

In the current theoretical framework, high energy lepton-hadron and
hadron-hadron interaction cross-sections $\sigma $, both in standard model
and in new physics processes, are related to calculable fundamental parton
interaction cross-sections $\widehat{\sigma }$ by the QCD factorization
theorems\cite{FacThm} as a sum of integrals convoluting the latter with
universal parton distribution functions. The parton distributions can, in
principle, be determined from analyzing a set of standard experiments ---
deep inelastic scattering (DIS), lepton-pair production (LPP), high-$p_{t
\text{ }}$ direct-photon production, W- and Z-production, high-$p_{t\text{ }%
} $ jet-production, heavy flavor production, ... etc. As both theory and
experiments have matured and grown in complexity and as the scope of the
standard experiments expands to various collider processes, the task of
proper and systematic global QCD analyses requires coordinated efforts of
both theorists and experimentalists familiar with details of both sides of
such an analysis. For this reason, the CTEQ Collaboration has been
developing the necessary tools for carrying out up-to-date global analyses
based on previous work of Duke-Owens\cite{DuOw} and Morfin-Tung\cite{MoTu}.

Previous global analyses were based on data from the SLAC-MIT, EMC, CDHSW,
and BCDMS deep inelastic scattering experiments; the E288, E605 lepton-pair
production experiments; and the WA70 direct photon production experiment.%
\cite{OwTu} Recently released NMC data\cite{NMC} on $F_2^n/F_2^p,%
\;F_2^p-F_2^n,$ and $F_2^{p,d}$ using a muon beam and CCFR data\cite{CCFR}
on $F_{2,3}^{Fe}$ using (anti-) neutrinos are expected to have a significant
impact on QCD global analyses because of their extended kinematic coverage
(particularly at small $x$), their high statistics and minimal systematic
errors. The precision of the current generation of DIS experiments now far
exceeds the size of next-to-leading order QCD contributions to these
processes; thus they probe the full complexity of QCD mixing effects between
quarks and gluons in a properly conducted QCD analysis. We find that these
new accurate DIS data, supplemented by constraints imposed by data on
lepton-pair and direct photon production, leads to substantially increased
sensitivity to the flavor content of the sea-quark distributions as well as
to the gluon distribution --- features conventionally thought to be beyond
the reach of analyses involving totally inclusive data.

To demonstrate the need for new comprehensive global analyses and the
potential for discovering new features of parton distributions in light of
the new data, we show in table I values of $\chi ^2$ per degree of freedom
obtained by comparison of old and new data with theoretical numbers obtained
{}from parton distributions widely used in the current literature: KMRS\cite
{KMRS}, MT-S1, MT-B2\cite{MoTu}, and the recent MRS D0\cite{MRS92} (which is
based on earlier
preliminary NMC and CCFR data). We note that all old sets do rather badly
with the new data. Only the MT-B2 set is close to being acceptable, but
still misses the CCFR data by 1.5 - 2 standard deviations. Even the MRS D0
set (which used the new data as input) does poorly, with most of its high $%
\chi ^2$ also coming from the CCFR data.%
\footnote{These large $\chi^2$ values are consistent with the original
MRS analysis
because the errors were enlarged in that analysis.}

\paragraph{General Description of New Analysis}

For the current analysis, we use the following sets of data: BCDMS $F_2^{\mu
p}\ \&\ F_2^{\mu d}$, NMC $F_2^{\mu p},\ F_2^{\mu d}\ \&\ F_2^{\mu
n}/F_2^{\mu p}$,%
\footnote{Although the ratio measurement is not totally independent from the
individual
$F_2^{\mu p}\ \&\ F_2^{\mu d}$, it is in fact more significant than the latter
since it involves
considerably smaller systematic errors and a somewhat extended kinematic range;
hence this data set is included upon the advice of some members of the NMC
collaboration.}%
\ CCFR $F_2^{\nu Fe}\ \&\ F_3^{\nu Fe}$ DIS structure functions, E605 LPP $%
d\sigma /dyd\tau $, and WA70, E706 \& UA6 direct photon $d\sigma /dydp_t$.
With kinematic cuts ($Q^2>4\ GeV^2$ in DIS and LPP, $W^2>12\ GeV^2$ in DIS,
and $p_t>4\ GeV$ in direct photon production), the total number of data
points used in most of our analysis is 917. We have tried higher $Q^2$-cuts
and found the results are not sensitive to the choice. For the very
high-statistics DIS experiments, statistical errors have been reduced to
such a low level that systematic errors often dominate the experimental
uncertainties. The systematic errors clearly cannot be neglected as has been
done in some past analyses. On the other hand, a complete treatment of these
errors, including all relevant correlations (as is usually done in
single-experiment analyses), would be (a) entirely impractical in the
context of a large-scale global analysis, and (b) uncertain in statistical
significance since such a diverse set of experimental data is involved.
Thus, we adopt the common practice of combining the statistical and
systematic errors in quadrature point-to-point. We have also tried to add
the errors linearly, and found similar conclusions (with, of course, reduced
absolute values of $\chi ^2$). The relative normalization of the various
data sets is allowed to vary around 1.0 during the fitting process with an
associated $\chi ^2$ included using the appropriate quoted experimental
normalization errors. Heavy target corrections are applied to the neutrino
structure functions to generate equivalent isoscalar structure functions for
nucleons. The correction factor is taken from measured ratios of structure
functions on light to heavy targets measured in muon scattering experiments.%
\cite{heavyT} In the crucial low-x region, the default correction factor
taken for this analysis is a conservative factor based on the assumption of
saturation of the shadowing effect for A $>$ 40. Since this correction is
significant compared to current experimental errors, we have also
investigated changes resulting from using a correction factor based on a fit
of shadowing data off He, C and Ca with subsequent extrapolation to Fe. The
differences will be noted later.

We perform least $\chi ^2$ fits using primarily the MINUIT program. All
results have been verified and investigated by a completely independent
program package developed by Duke and Owens.\cite{DuOw} We use the overall $%
\chi ^2$ as well as individual $\chi ^2$'s for each experiment as measures
of the {\em relative} ``goodness of fit'', but do not attach strict
statistical significance to the absolute values of the $\chi ^2$'s for
reasons mentioned above. For quantitative next-to-leading order (NLO) QCD
global analyses based on data of high accuracy, one must take into account a
number of non-trivial theoretical and experimental considerations. These
issues have been described in detail in several reviews,\cite{OwTu},\cite
{TuDallas},\cite{pdfwks} hence will not be discussed here.

Recent experiments have called into question a number of traditional
assumptions about the unknown input parton distributions (at some fixed
scale $Q_0^2$) such as SU(3) or SU(2) flavor symmetry for the sea-quarks.
Since the diversity of data sets used in the analysis and the measurement
accuracy have increased sufficiently to provide stringent constraints on the
theoretical unknowns, we find it desirable to try fits which are free from
most conventional prejudices and compare the results with those obtained
with more traditional assumptions. This comparison turns out to be very
revealing.

For most of our analyses, the QCD evolution kernel, the effective coupling,
and all hard scattering cross-sections are taken to be in the \Msbar   \
renormalization scheme; hence the extracted parton distributions are, by
definition, in the \Msbar   \ scheme. For the convenience of certain
applications, we also obtained parton distribution functions defined in the
DIS scheme. These distributions are generated by independent fits to the
same data.%
\footnote{This is a more reliable procedure then by transforming from the
\Msbar\ sets since the ``NLO'' term in the conversion formula can,
under some circumstances, be comparable in size to the ``LO'' term, thus
vitiating the
perturbative nature of the transformation.\Ref{MoTu}.} Likewise, we generate
representative leading-order fits for applications which use LO hard matrix
elements. For concreteness, in the following description of our results, we
shall focus on the \Msbar  \ distributions.

\paragraph{Description of Results}

The most notable fact of this new global analysis is the extraordinary
quantitative agreement of the NLO-QCD parton framework with the very high
statistics DIS experiments over the entire kinematic range these experiments
cover and the consistency of this framework with all available experiments
in lepton pair and direct photon production as well. In the least
constrained fits, the overall $\chi ^2$ is typically around $860$ for about
$880$ degrees of freedom (917 data points with $30\sim 35$ parton shape and
relative experimental normalization fitting parameters); and the $\chi ^2$
for the individual experiments are all comparable to the respective number
of points --- a remarkable result considering the diversity of processes,
the kinematic range covered, and the accuracy of many of the experiments.
Other fits with certain restrictions (such as SU(2) or SU(3) symmetric sea)
have overall $\chi ^2$ around $910$ or higher. Although the relative values
of $\chi ^2$ are considerably larger than the best fit, even these
alternatives can represent reasonable fits when taken by themselves. The
plots of these fits against data (not included in this letter) give the
impression of good agreement in most cases and only very careful examination
will reveal the differences. The merit of these possible solutions can be
better judged by examining details of the parton distributions obtained and
of the comparison with the various data sets used, as will be discussed
later.

In general we find two common features of the new fits:

\noindent
(i) The small-$x$ $(\sim 0.01-0.08)$ data from NMC and CCFR require
substantially increased sea-quark distributions in this region. We found
this also produces an indirect effect on the shape of the gluon distribution
such that the momentum fraction carried by the gluon at $Q=2\ $GeV is
reduced from the previously accepted value $\sim \,0.45$ to around 0.42.
This fraction is partially recovered after evolution to larger $Q\,$. (Note
that in NLO QCD the gluon momentum fraction is factorization scheme
dependent (i.e. different in the \Msbar    \ and DIS schemes); and it is not
exactly tied to the integral of the neutrino structure function $F_2$ as in
leading order.)

\noindent
(ii) As a free-parameter in the global fit, the NLO $\Lambda
_{QCD}(5\,flavors)$ is found to be typically $150\ $to $160\ $MeV in these
new fits%
\footnote{This result supersedes
a preliminary CTEQ value reported at the 1992 International High Energy
Conference at
Dallas\cite{TuDallas} which is inaccurate due to a computer error.} ---
consistent with typical values [$\Lambda _{QCD}(4fl)=220\ $to $240\ $MeV ]
found in direct $d\ln F_2/d\ln Q^2$ analyses\cite{qcdlamExp} and in previous
global analyses.\cite{MoTu},\cite{MRS90} Fits with $\Lambda _{QCD}(5fl)$
fixed at $220\ $MeV (corresponding to a higher value of $\alpha _s$ favored
by some LEP measurements) have substantially higher $\chi ^2$ --- by about
30.

In Table II we present a set of representative fits with different
characteristics and list their overall $\chi ^2$ as well as the $\chi ^2$
per data points for the various data sets. The best fit in the \Msbar    \
scheme, with the least constraints, is designated CTEQ1M.%
\footnote{We adopt the following label scheme for our parton distributions:
CTEQnSX,
where n is a version number; S designates the factorization scheme -- M for
\Msbar,
D for \dis, and L for \lo; X is absent for the best fits, otherwise it
distinguishes the alternative sets within a given scheme.  Parametrizations
of these distributions can be obtained
by sending a request to Botts@MSUPA (bitnet) or Botts@MSUPA.PA.MSU.EDU
(internet).}
For this set, the input distribution functions at $Q_0$ ($=2$ GeV) for all
the flavors are taken to be of the form: $%
f(x,Q_0)=A_0x^{A_1}(1-x)^{A_2}(1+A_3x^{A_4})$. The coefficients $A_i$ are
subjected to minimal restrictions in addition to the quark-number and
momentum sum rules. Although, it may appear that there are many parameters
to be determined, it turns out that the wealth of data provides surprisingly
stable results for the best fit irrespective of where the starting point is
taken to be in the parameter space. A glance across the first row of Table
II reveals that all data sets are uniformly well-fit%
\footnote{The only apparently ``high'' $\chi^2$ associated with the NMC
hydrogen date
is due to differences of the NMC and BCDMS data around the $x=0.225$ bin
where the more
abundant and more accurate BCDMS data dominate the fit.}, indicating a good
deal of consistency. The most conspicuous feature of the CTEQ1M
distributions is the rather large flavor dependence of the sea quarks ---
the $\bar u$, $\bar d$, and $\bar s$ distributions all differ substantially
in shape. See Fig.1. In addition, because of the significantly increased $%
s(x)$ in the small-$x$ region, the normalization of the strange quark, as
measured by the ratio of momentum fractions (integrated over the full $x$%
-range $(0,1)$) ${\kappa =}2s/(\bar u+\bar d)$, is around $0.9$ instead of
the often quoted value of 0.5.\cite{ccfrKappa} Since these results appear to
be unusual, we will discuss them in some detail after we describe the other
fits in order to provide a basis for comparison.

CTEQ1D and CTEQ1L are DIS and LO{\ distributions with the same assumptions
as CTEQ1M. The DIS distribution gives equally good fit to data as the
\Msbar  \   one, just as one expects -- a consistent change in factorization
scheme should not change the physics. The LO fit has a substantially higher $%
\chi ^2$} (by $\sim 50$) and associated normalization correction factor for
the LPP data, confirming the need of the NLO formalism for precision QCD
analyses. CTEQ1MS is a set of distributions with a ``singular'' gluon
behaving like $x^{-0.5}$ at $Q_0;$ its $\chi ^2$, comparable to the best
fits, indicates that current data allow a range of different extrapolations
to smaller $x$ beyond 0.01.  And CTEQ1ML is a set obtained with $\Lambda
(5fl)$ fixed at $220$ MeV (the ``LEP event shape
value''). All these distributions give
reasonable fits to data with different characteristics (and some with
associated limitations, such as the LO one). We shall not discuss them
further in this short report.

The row labeled {M2 represents a fit assuming a SU(2) flavor symmetric sea.
The overall $\chi ^2$ is higher than that of CTEQ1M by 50,  all data sets
are uniformly well-fit as for the latter. However, it is associated with a
normalization of the strange quark corresponding to $\kappa =1.41$ -- too
high a value from the physics point of view. } {In order to investigate the
sensitivity of the combined data set to the strange quark normalization and
shape, several fits with $\kappa $ fixed at 0.5 (according to conventional
wisdom) are tried. The last row of Table II, labeled K5, represents the best
fit one can get with this restriction. The overall $\chi ^2$ increased
dramatically by 88. If one requires the shape of }$s(x,Q_0)$ to be the same
as the non-strange sea distributions, {{{{{the $\chi ^2$ also increases very
substantially. Thus, contrary to conventional expectations, we found {\em %
this data set \ (consisting of only totally inclusive quantities) is
sensitive to $\kappa $ and to the shape of $s(x,Q_0)$, and it strongly
prefers $\kappa \approx 0.9$ or higher}. }}}}}

To trace the origin of this surprising result, we note the principal
difference between the CTEQ1M and the K5-fits in Table II is the increased $%
\chi ^2$ values of the CCFR $F_2^\nu $ and the NMC $F_2^{\mu d}$ data sets.
In leading order QCD, $F_2^{\nu d}$ (converted from $F_2^{\nu Fe}$ using a
heavy-target correction factor) is given by the straight sum of all quark
flavors whereas $F_2^{\mu d}$ corresponds to the same sum weighted by the
squared charges. One can therefore get a direct handle on the strange quark
distribution by examining the equality $\frac 56F_2^{\nu d}-3F_2^{\mu
d}=xs(x,Q)\;+\;O(\alpha _s),$ valid at small $Q$ where the charm
distribution is small. Fig.2 shows a plot of the quantity on the left-hand
side of this equation at $Q^2=5\;$GeV$^2$ using data from NMC and CCFR, and
compares it with the strange quark distributions of CTEQ1M, K5, and MRS D0 at
the same $Q^2$. The size of the experimental errors are superimposed on the $%
\frac 56F_2^{\nu d}-3F_2^{\mu d}$ curve. It is clear from this plot that $%
\kappa \simeq 0.5$ parton distribution sets are not compatible with the new
data; and that both the size and the shape of $s(x,Q)$ in the best fit are
driven directly by data. How can this conclusion be reconciled with the
conventional low value of $\kappa $ obtained from leading order parton model
analysis of dimuon production data?\cite{ccfrKappa} First, we note that our $%
\kappa $ is defined at a fixed $Q^2$ (=4 GeV$^2$) and is integrated over the
full $x$-range $(0,1)$; whereas the $\kappa $ determined from the dimuon
experiments use data from a wide range of $Q^2$ extending from less than 1
GeV$^2$ for the lowest $x$ values to well over 100 GeV$^2$ at moderate values
of $x$. The two quantities are not the same as $\kappa $ is $Q$-dependent for
a non-SU(3) symmetric sea. Secondly, from Fig.1, we see that for $x\,>0.1$
our $s(x,Q)$ actually becomes considerably smaller then $\bar u$ and $\bar d$
as one expected. If $\kappa (x_{\min })$ is defined as the ratio of momentum
fractions integrated from some $x_{\min }$ to 1, then $\kappa (0.1)\approx
0.5;$ it increases to $\kappa (0.01)\approx 0.85$ and eventually to $\approx
0.9$ in the theoretical limit $x_{\min }\rightarrow 0$. Thus, the comparison
with the dimuon experiment is also dependent on the effective $x$-range and
the extrapolations used in going beyond that range. Finally, as has been
pointed out,\cite{AOT} existing LO dimuon analyses are likely to be
unreliable at small $x$ due to the hitherto neglected contribution from the
gluon-fusion mechanism which is nominally next-to-leading order but
physically significant.

The question of SU(2) symmetry of the non-strange sea can be investigated by
comparing the two fits CTEQ1M (non-symmetric, $\kappa =0.9$) and M2
(symmetric, $\kappa =1.4$) in Table II. The increase in the overall $\chi ^2$
of 52 mainly comes from the CCFR $F_{2,3}$ and E605 data sets.  We know that
the $pN$ LPP cross-section should be sensitive to $\bar u$ because of the
dominance of $u\bar u$ scattering.%
\footnote{The CTEQ analysis uses the full E605 data set on $d\sigma /dyd\tau $
(120 points); whereas, it appears from the published literature that, the MRS
analyses use only the
integrated $d\sigma /d\tau $ data for consistency checks\cite{MRS90}.} When
restrictions are imposed on the flavor dependence of the sea, it is the
interplay between the CCFR, NMC and E605 data sets which causes the $\chi ^2$
to increase. While M2 has an (arguably) acceptable $\chi ^2$, it does have
an abnormally large strange quark component. If one restricts the size of $%
s(x,Q_0)$, say $\kappa \leq 1.0$, any solution with an SU(2) symmetric sea
will have an unacceptable high $\chi ^2$. We also note, in the preferred
set, CTEQ1M, the $\bar u$ and $\bar d$ distributions \ cross each other
around $x\sim 0.06$. This behavior contrasts with that of MRS D0 where $(\bar
d-\bar u)$ is chosen to be positive definite. We have found that the
behavior shown is preferred for all the fits we have tried whenever the sign
of this difference is left  free.%
\footnote{This may be partially responsible for the very high $\chi ^2$ values
on the
CCFR data using the MRS D0 distributions (cf. Table I).} In this connection,
we mention there is some tantalizing experimental evidence for this behavior
of $(\bar d-\bar u)$ from E772 at Fermilab.\cite{e772} We know of no reason
why this quantity need be positive definite.

Details on the above mentioned fits will be described in a full-length paper.

\paragraph{Uncertainties and Challenges}

There are uncertainties associated with this analysis some of which can be
addressed. Concerning the heavy target correction in the small-$x$ region,
we found the use of alternative shadowing correction schemes (cf. Sec.2)
leads to larger upward corrections to $F_2^{\nu d}$, hence an even bigger
strange-sea. One may also consider the effect of shadowing inside deuterium
on relating deuterium to nucleon structure functions used in the analysis.
Recent theoretical studies indicate that this effect is less than 2-3\% at
the smallest-$x$ value covered here.\cite{Strikman} This is small compared
to the uncertainties of the heavy-target correction. Detailed work can be
done to quantify both these uncertainties. We also note, current analyses of
DIS neutrino structure functions use the NLO QCD formalism for massless
quarks, including the small but non-negligible $s\rightarrow c$
contribution. The effect of the charm quark mass is taken into account by
applying a ``charm-threshold correction'' to the data, not at the
theoretical end. This is not fully satisfactory because the applied
``correction'' is, at best, a leading order one. Since the complete NLO
theory for this transition now exists,\cite{NLOdimu} the correct procedure
is to include this in the theoretical expression and to compare with the
{\em uncorrected} physical structure functions. We do not, however, expect
either of these possible improvements to affect the main features of our
results since, as we have demonstrated, they are driven rather directly by
the new data. The only alternatives to the large strange sea at small-x
would be: (i) there are additional unspecified
systematic errors associated with the two major experiments at small-x; and
(ii) the theoretical corrections due to charm-mass effects are much larger
than expected and turn out to account for the bulk of the observed
difference $\frac 56F_2^{\nu d}-3F_2^{\mu d}$ in place of $xs(x,Q)$. The
last possibility is under investigation.

Hadron collider data on W-, Z-production, lepton pair production,
direct-photon production, jet production, and heavy flavor production are
providing useful tests of the conventional QCD calculations. The expected
increase in integrated luminosity by an order of magnitude at the Tevatron
will make these measurements important sources of quantitative information
on parton distributions. The new parton distributions obtained in our global
analysis lead to increased cross-section for all the above mentioned
collider processes in the regions dominated by small-$x$ partons (low mass
lepton pair, low $p_t$ direct photon, and heavy flavor production). This is
gratifying since earlier parton distributions tend to give smaller
cross-sections on all these processes when compared to preliminary CDF data.
Input from new results obtained in the current run at the Tevatron will help
probe the very small-$x$ region; and will measure, in addition, different
combinations of parton distributions than currently available. Given the
tight constraints on these functions already found in this analysis, the new
input will certainly provide significant quantitative tests of the
consistency of the QCD parton framework and lead to even better
determination of the parton distributions. We also, of course, eagerly
anticipate results from the HERA ep collider\cite{hera} which will provide a
wealth of information on structure functions and other observables sensitive
to parton distributions in a widely expanded kinematic range.

\paragraph{Acknowledgment}

We would like to thank our colleagues on the CTEQ Collaboration for
consultation and for encouragement. We would also like to thank Alan Martin
and Dick Roberts for useful communications.

\rule{0mm}{1cm} \\
\noindent {\em Note Added in Proof:}

Since this paper was submitted, a new leading-order QCD analysis of dimuon
events in neutrino scattering has been published by the CCFR group (PRL 70,
134 (1993)), reaffirming a smaller strange sea than the non-strange one.  The
difference in the measured $s(x,Q)$, compared to that obtained from our global
analysis, lies mainly in the small $x$ region.  We have been informed by the
CCFR collaboration that preliminary results from a next-to-leading order
analysis of the same data do not lead to substantial changes in these results.
If this measurement of the strange sea is correct, then the difference of the
neutrino and muon $F_2$ structure functions shown in Fig. 2 cannot be fully
attributed to the strange quark. This suggests either additional sources of
theoretical uncertainties so far neglected in all global analyses, or an
incompatibility of the $F_2$ measurements of the CCFR and NMC experiments
within the stated errors in the very small $x$ region.  To see how changes in
experimental systematic errors might affect the situation, we have performed
new fits fixing $\kappa = 0.5$ and increasing the experimental systematic
error in the smallest-$x$ bins by a multiplicative factor.  Reasonable fits
with $\chi^2$ in-between the best CTEQ fits and the "$\kappa$5 fit" (cf. Table
II) are obtained with a factor of order 2 for either of the experiments.
Aside from readjustments among the sea-quark flavors, the shapes of the other
parton distributions remain quite stable. Details will be described elsewhere.

It is important to emphasize that the question about the strange quark should
not obscure the main results of this paper: the determination of a new
generation of parton distributions based on recent highly precise experimental
measurements and phenomenological advances. The progress made is evident in
the $\chi^2$ comparison between Tables I and II. The new degree of accuracy is
also responsible for bringing the potential conflict between the new $F_2$
measurements and the dimuon results to light.  To resolve the current puzzle
we need to reevaluate all theoretical uncertainties; improve the treatment of
systematic errors in the global analysis; and, finally, include the dimuon
data in the analysis. These efforts are underway. At the same time, we urge
the key experiments to re-examine their data in the small-$x$ region to see if
our observed discrepancy persists under closer scrutiny.

We thank various members of the CCFR Collaboration, particularly Sanjib Mishra
and Michael Shaevitz for in-depth discussions on the strange quark issue.

\pagebreak

\oddsidemargin=0.0in
\topmargin=-.3in
\textwidth=6.5in
\baselineskip=14pt

{\small{
\begin{tabular}{|c|c||ccccccccc|}
\hline
& Total $\chi^2$ & ccf$F_2$ & ccf$F_3$ & nmcH & nmcD & nmcR & bcdH
& bcdD & E605 & Dir $\gamma$ \\ \hline
KMRS B0 &
3954 & 22.16 & 3.34 & 6.11 & 8.46 & 1.29 & 1.25 & 1.40 & 1.20 & 1.02 \\
MT S1 &
3786 & 21.56 & 4.32 & 4.43 & 6.38 & 1.81 & 1.57 & 1.44 & 1.00 & 1.77 \\
MT B2 &
1415 &  3.10 & 3.14 & 2.21 & 1.41 & 1.27 & 0.76 & 1.05 & 0.90 & 2.46 \\
\hline \hline
MRS D0 &
1643 & 5.31 & 3.71 & 1.47 & 1.24 & 1.27 & 1.17 & 1.21 & 1.16 & 0.96 \\
\hline
\end{tabular}
}}
\hspace{1em}

\begin{flushleft}
{\bf Table I:}  $\chi^2$ of some widely used parton distributions applied
to current experimental data.  The first column gives the total
$\chi^2$ for $917$ data points (all experiments).  The other columns
give the $\chi^2$ per data point for the individual data sets:
$CCFR$ $F_2$ and $F_3$; $NMC$ $H$, $D$, and $F_2^n / F_2^p$;
$BCDMS$ $H$ and $D$; $E605$; and the combined direct photon
production data ($WA70$, $E706$ and $UA6$).
\end{flushleft}

\hspace{4em}

{\small{
\begin{tabular}{|c|c||ccccccccc|}
\hline
& Total $\chi^2$ & ccf$F_2$ & ccf$F_3$ & nmcH & nmcD & nmcR & bcdH
& bcdD & E605 & Dir $\gamma$ \\ \hline
\# pts. & 917 & 77 & 77 & 83 & 83 & 89 & 168 & 156 & 119 & 47 \\ \hline
CTEQ1M   & 860 &0.68 &0.71 & 1.25 &0.90 & 1.34 &0.72 & 1.12 &0.87 &0.73 \\
CTEQ1D   & 861 &0.74 &0.90 & 1.26 &0.91 & 1.30 &0.65 & 1.07 &0.83 &0.70 \\
CTEQ1MS  & 861 &0.71 &0.79 & 1.26 &0.89 & 1.25 &0.71 & 1.10 &0.87 &0.74 \\
\hline \hline
CTEQ1ML  & 892 &0.78 &0.80 & 1.24 &0.86 & 1.28 &0.82 & 1.18 &0.83 &0.75 \\
CTEQ1L   & 914 &0.89 &0.86 & 1.03 &1.07 & 1.50 &0.68 & 0.99 &1.15 &0.73 \\
\hline \hline
M2 & 912 & 0.81 &0.96 & 1.28 & 0.87 & 1.30 & 0.76 & 1.11 & 0.99 & 0.89 \\
$\kappa$5
   & 948 &0.96 & 0.87 & 1.26 & 1.00 & 1.28 & 0.72 & 1.14 & 1.09 & 0.86 \\
\hline
\end{tabular}
}}
\begin{flushleft}
{\bf Table II:}  $\chi^2$ of various $CTEQ$ global fits to current data.
Column labels are the same as in Table I.  The number of fitting
parameters (parton distribution shape and relative experimental
normalization) is of the order of $30$ to $35$; resulting in about
$880$ degrees of freedom.
\end{flushleft}

\pagebreak

\begin{center}
{\bf Figure Captions}
\end{center}

\paragraph{Fig. 1}

: Sea-quark distributions. The flavor labels on the left are for CTEQ1M;
those on the right for MRS\_D0.

\paragraph{Fig. 2}

: Comparison of experimentally measured $\frac 56F_2^{\nu d}-3F_2^{\mu d}$
with $xs(x, Q_0)$ from two of our fits, CTEQ1M and K5, and from MRS\_D0.
Errors of the two experiments are added in quadrature and superimposed on
the relevant curve to indicate the size of the experimental uncertainty.


\begin{thebibliography}{99}
\bibitem{heavyT}  We used an emperical formula based on NMC measurements at
small $x$ and SLAC measurements at large $x$. (Private communication from A.
Bruell (NMC).
\end{thebibliography}
\end{document}